# Transparently Capturing Request Execution Path for Anomaly Detection

Yong Yang, *Student Member*, *IEEE*, Long Wang, *Senior Member*, *IEEE*, Jing Gu, Ying Li, *Member*, *IEEE*

**Abstract**— With the increasing scale and complexity of cloud systems and big data analytics platforms, it is becoming more and more challenging to understand and diagnose the processing of a service request in such distributed platforms. One way that helps to deal with this problem is to capture the complete end-to-end execution path of service requests among all involved components accurately. This paper presents REPTrace, a generic methodology for capturing such execution paths in a transparent fashion. We analyze a comprehensive list of execution scenarios, and propose principles and algorithms for generating the end-to-end request execution path for all the scenarios. Moreover, this paper presents an anomaly detection approach exploiting request execution paths to detect anomalies of the execution during request processing. The experiments on four popular distributed platforms with different workloads show that REPTrace can transparently capture the accurate request execution path with reasonable latency and negligible network overhead. Fault injection experiments show that execution anomalies are detected with high recall (96%).

✦

## 1 INTRODUCTION

Distributed platforms like cloud platforms (e.g. IBM Cloud, Amazon AWS, OpenStack and SoftLayer) and analytics platforms (e.g. IBM Watson Health Cloud [32], Spark and Hadoop) are composed of numerous services and components in multiple levels to support various applications. Expertise on such platforms and hosted services is critical for the enterprises of platform providers such as IBM, Microsoft, Amazon, and Google, to maintain and improve the platforms and services.

However, it is challenging for the platform provider enterprises to obtain and keep the up-to-date expertise on the platforms and hosted services/applications because, i) new services and applications continue to onboard the platforms; ii) softwares of such platforms and their hosted services/applications have been developed by different vendors or teams and may be owned by different parties, and source code of them may be unavailable; iii) the system softwares and the services are under continuous and rapid updates (e.g. OpenStack is released every 6 months [29]), iv) there is frequent flow of personnel in operation/development teams (this makes the knowledge transfer a headache for most large providers of computing platforms). Actually, it is reported that up to 60% of the software engineering effort is spent on understanding the software system [30].

Drawing message flows of individual request's processing among the components of a distributed platform was demonstrated to be an effective way to understand and analyze behavior of cloud platforms. For instance, the analysis of OpenStack's release evolvements [28] presented manually drawn message flows among OpenStack's components for identifying behavior changes in different OpenStack releases.

Manual construction of message flows or other capturings of system behaviors is inefficient towards the goal of obtaining/keeping up-to-date understanding/expertise on such a platform and hosted services. Moreover, we intend to *capture the complete end-to-end execution path of processing an individual request among all involved components of the platform*. We believe the complete path conveys a holistic view of the platform's behavior during the processing of the request, and this holistic view, with in-depth details, boosts the up-to-date understanding/knowledge and even diagnosis of request processing behavior (as demonstrated in our work in [25]), especially in our specific context where the softwares of the platform and hosted services/applications are from different vendors/teams/owners and under constant changes, relevant source code is unavailable or rapidly evolving, and the platforms have frequent flow of personnel. *This capturing of the end-to-end path in our specific context truly differentiates our work from previous ones*, which are discussed below, and is attractive to enterprises providing such heterogeneous platforms like IBM Watson Health Cloud.

A number of technologies in current literature try to tackle a similar problem (end-to-end tracing). Stardust [15] is designed to trace a specific distributed storage system and requires manual modification of the system source code. Xtrace [13] extends common network protocols to support request identifier propagation by means of special hardware and application support of the protocol extension. Google's Dapper [6] instruments Google's internal homogeneous control flow/RPC libraries to do the tracing.

---

- *Yong Yang is with Peking University, Beijing, China, 100871. E-mail: yang.yong@pku.edu.cn*
- *Long Wang is with IBM Waston, New York, USA, 10514. E-mail: wanglo@us.ibm.com*
- *Jing Gu is with Peking University, Beijing, China, 100871. E-mail: gu.jing@pku.edu.cn*
- *Ying Li is with Peking University, Beijing, China, 100871. E-mail: li.ying@pku.edu.cn*

It requires all traced components use the instrumented internal libraries. Pinpoint [11] instruments J2EE libraries to accurately correlate network communication events by adding request identifier in HTTP headers. The tracing tools of Pivot [3], Magpie [12], HDFS HTrace [17] and Facebook's Canopy [1] are similar to Dapper and Pinpoint in that they instrument specific middleware or vendor custom libraries and the instrumentations enable the tracing capabilities. Other related work e.g. statistical approaches are discussed in Section 6.

What we desire is a mechanism of capturing the complete request execution path for distributed platforms bearing the four challenges above. The existing technologies above do not apply in our target situations where the challenges i), ii) and iii) are combined. A project close to our work is vPath [7]. This project traces communications of distributed systems/applications by pairing each network message' sending (by the sender thread $S$) and receiving (by the receiver thread $R$) via the TCP/IP socket and connection information (e.g. source IP, source port, destination IP, destination port, source thread ID and destination thread ID). Because vPath relies on pairing of TCP/IP socket and connection information for identifying causality, this makes vPath work only in one type of communication/thread pattern: a) $S$ sends message to $R$ via connection $c$ and blocks itself; b) if $R$ replies to $S$ via connection $c$ then $S$ resumes its execution, or if $R$ sends message to another thread $T$ via another connection $d$, $R$ blocks itself until it receives reply from $T$ via $d$. For any other communication pattern or thread pattern, vPath will fail (as the vPath authors stated in [7] and our experiments also showed). These other patterns include asynchronous communication, thread forking or thread handover during request processing, connectionless communication (UDP), connection pool, event driven and thread pool.

Our work was motivated by vPath. We intend to build a general mechanism that traces distributed platforms' processing of individual requests covering most, if not all, execution scenarios, after we found vPath did not work in our platforms due to the pattern constraint. We also intend to capture the complete end-to-end path while vPath reports path fragments for a request because the patterns beyond what vPath supports are common.

In this paper a Request Execution Path (REP as acronym) is defined as the complete path of processing a specific service/job request. Specifically, the REP 1) covers all of the executions of the given distributed platform's components in processing a specific request (at a given granularity such as library/system calls), including executions of those components' processes and threads as well as the communications among them during the processing; 2) identifies all these executions, and links them together according to the accurate causal relationships among them to form an integral unseparated view of the platform's processing.

We propose Request Execution Path Trace (REPTrace) to capture the REP. REPTrace intercepts runtime events such as common library/system calls at the operating system level (either within a physical machine, VM, or container). Principles and algorithms are proposed for identifying the relationships among the events. Then REPTrace stitches all the events of one request's processing together using the identified relationships, and produces the complete REP. To satisfy the challenges in our problem context above, we i) perform comprehensive analysis of execution scenarios about how distributed systems process requests, ii) modify network messages with extra information for pairing the sending and receiving of the same message, iii) maintain the context of each request's processing to differentiate it from other concurrent requests, and iv) devise algorithms to combine these three items and maneuver a set of relevant library/system calls for tracing the analyzed execution scenarios.

As a demonstration of REPtrace usage, this paper also presents an anomaly detection approach that leverages REPs to detect anomalies during processing of a request. Besides anomaly detection, REPTrace helps understand behavior of complicated distributed platforms [25]. In summary, this paper makes the following contributions:

• We analyze execution scenarios and event relationships in these scenarios for request processing of distributed platforms and identify event relationships in all the execution scenarios. This analysis is the basis for REPTrace.

• We propose REPTrace for constructing complete REPs without code instrumentation, and implement REPTrace on Linux systems. Novel algorithms are devised for the REP construction and no specific communication or thread patterns are assumed in REPTrace. As far as we know, there is no prior art of distributed tracing that tries to address the comprehensive communication and execution scenarios in the request execution path as we handle.

• A novel anomaly detection approach based on REPTrace is proposed to detect execution anomalies by discovering deviations of the execution paths from normal ones.

• We evaluate REPTrace and the anomaly detection method with extensive experiments in terms of performance overhead, anomaly detection precision, recall and F1-measure on different distributed platforms. Our experimental results show that with a 4.5% execution latency overhead the REPTrace-based anomaly detection approach detects functional anomalies with 93% precision and 96% recall, and detects performance anomalies with 74% precision and 74% recall.

## 2 OVERVIEW OF REP CONSTRUCTION

If we look into the execution of a request processing, we see the execution consists of the following sequence: ... - *event – thread execution – event – thread execution* - ... Here *event* indicates a system call (or a LIBC call encapsulating a system call) that is of our interest, and *thread execution* is one thread's continuous execution between one event and its successive event. After removing *thread execution* from the sequence, the execution process is then denoted as the sequence ... - *event – event* - .... In this paper we use such sequences of events to denote the execution of request processing. Suppose an event $e1$ has its immediate successive event $e2$ in the sequence; we call $e1$ *parent* of $e2$ and $e2$ *child* of $e1$, and their relationship is denoted as $e1->e2$ (*parent->child*). Note that one event may have multiple parents when it is in multiple sequences (we will elaborate on this in Section 2.3.2).

## 2.1 Analysis of Execution Scenarios

The following is a comprehensive list of scenarios (illustrated in Fig. 1) that may be involved during service/job request processing based on our experience and analysis of real-world IT paltforms. We believe the comprehensive list covers most, if not all, of the common execution scenarios in request processing. We do this analysis as we intend to support scenarios as many as possible rather than limited patterns only.

   a) Continuous execution within a thread;
   b) The current thread creates another thread and passes the handling of the request to the new thread. The current thread may stop processing (e.g. sleep), or continue processing this request;
   c) The current process forks a process, and passes the handling of the request to the new process; The current process may stop or continue processing this request;
   d) The current process/thread sends a message to another process/thread, which may be on the same machine or a different machine. Then the latter process/thread begins the processing of the request. Network communication and other similar mechanisms, like pipe, are covered in this scenario; The current process/thread may stop or continue processing this request;
   e) The current process/thread synchronizes the processing of the request with another existing process/thread using certain IPC (Inter-Process Communication) mechanism such as process wait, thread join, signal, lock/unlock, semaphore, etc.;
   f) The current process/thread saves the request (or its intermediate state) in a message queue. Then a different process/thread picks up the request (or the intermediate state) from the message queue and begins processing;
   g) The current process/thread passes the handling of the request to another existing process/thread using shared memory, shared variables, or mapped device.

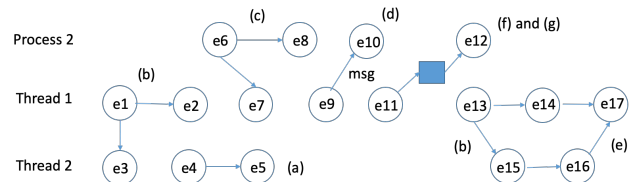

Fig. 1. Illustration of different categories of execution scenarios (note these are not parts of the same complete REP)

## 2.2 Event Relationships

The relationships between any parent event and its child event in the listed execution scenarios are summarized as the following types.

   1. *Temporal successiveness within the same thread (TCR)* in scenario *a* and other scenarios (e.g. in asynchronous messaging mode a thread continues processing of the request after sending message). Examples are *e4->e5, e1->e2, e6->e8, e13->e14->e17, e15->e16* in Fig. 1. Note that only those events that participate processing of the current request are considered, and the events of the same thread's processing of other requests in concurrency should be excluded for the REP generation of this request.
   2. *Causality of thread creation and process creation (ToPCR)* in scenarios *b* and *c* (*e1->e3, e6->e7, e13->e15*). In this type of relationship, the first event in the new thread/process is the child event of the thread-creation/process-creation event in the parent thread/process.
   3. *Causality of communication between threads/processes (COMR)* in scenario *d* (*e9->e10*). The communication events can be classified into two types, message sending type and message receiving type, which are respectively referred as *send* and *recv* hereafter. The *send* event in the sender thread is the parent of the corresponding *recv* event in the receiver thread.
   4. *Causality of synchronization among threads (SYNR)* in scenario *e* (*e16->e17* where *pthread_join* is the synchronization method). Depending on the synchronization mechanism, either the event that performs the synchronization is the parent of the first event in the target thread after it receives the synchronization signal, or the event that causes the syschronization signal in the target thread is the parent of the event that happens immediately after dealing with the syschroniztion signal in another thread.
   5. *Causality via data dependency between threads/processes (DDR)* in scenarios *f* and *g* (*e11->e12*). The data dependency is *read-after-write* dependency between threads/processes. The event that writes data in one thread is the parent of the event that reads the written value in another thread. If the write/read of the data is not via a system call but via direct memory access (e.g. shared memory, shared variable), the event that is immediately before the *write* is the parent event, and the event that is immediately after the *read* is the child event.

The event relationship types of 1~4 (scenarios *a~e*) are about manipulations or interactions of system control objects, i.e. thread, process, signal, communication, etc.; they indicate control flow of the request processing among threads/processes. The type of 5 (scenarios *f* and *g*) indicates data flow of the request processing among threads/processes.

## 2.3 Construction of REP

REP construction consists of two steps: generation of events and linking of events. For simplicity our description of REP construction here assumes that events are all generated in the first step, and then the linking is performed over the entire set of events in the second step. But in reality, the two steps on different events are performed in parallel, i.e. the linkings of certain events are being performed while certain other events are generated.

### 2.3.1 Event Generation

We intercept system calls of interest, or LIBC calls that encapsulate those system calls, and generate events during the interception. Table 1 lists the categories of calls that should be intercepted for event generation. According to the discussions above, the calls we are interested in are mostly of four categories: thread manipulation, process manipulation, network communication and synchronization. Certain other calls are also intercepted.

4TABLE 1: LIBC CALLS INTERCEPTED FOR EVENT GENERATION

| Category | Examples |
|---|---|
| Thread manipulation | pthread_create, pthread_join, pthread_self, pthread_detach, pthread_cancel |
| Process manipulation | fork, vfork, exec |
| Network communication | send, recv, write, read, sendmsg |
| Synchronization | wait, waitpid, pthread_join, signal |
| Other | open. close, malloc, syscall |

When a system call is intercepted by REPTrace, the original system call is encapsulated and invoked, both the parameters and results of the call are captured, and then a trace event for the system call calling event is generated. The trace event has common attributes that exist for categories of events, as well as attributes specific to the event's category. Common attributes include *event ID, call info, thread info, process info, node info, local time info, event linking info*. Here *call info* includes the name of the intercepted system/LIBC call as well as the parameters and return value of the call; *thread info, process info* and *node info* include information about the thread, process, and machine node where the interception occurs; *event linking info* includes attributes our algorithm creates or uses to link events and form REP, and is discussed in the Section below. The event-category specific attributes include IP address, port number, etc. Every system call calling event has a corresponding trace event (we use event and trace event interchangeably hereafter).

```
Algorithm 1 Event_Linking
Input: trace events set S
Output: parent[e] for each event in S
1.  For each event e in S :
2.     parent[e] = GetParent( e )
3.     If parent[e].call_id in [ thread_join, wait, waitpid ] : // type 4
4.        assign parents to e based on the call's specific semantics
5.
6.  GetParent( e ) :
7.     If e is of recv event and e.MSG_ID not empty : // type 3
8.        find event p in S for p.MSG_ID = e.MSG_ID
9.        Return p
10.    else
11.       find event p in S for p.MSG_CTX_ID = e.MSG_CTX_ID  // type 1
                and p.thread_id = e.thread_id
                and p.time_info is closest before e
12.    If p is found :
13.       Return p
14.    /* this means e is the first event of its thread/process */
15.    find event p in S for p.MSG_CTX_ID = e.MSG_CTX_ID // type 2
                and p.call_id = thread_create
                and p.return_value = e.thread_id
16.    If p is found :
17.       Return p
18.    /* this mean E is the first event of it's process */
19.    find event p in S for p.MSG_CTX_ID = e.MSG_CTX_ID // type 2
                and p.call_id in [fork, vfork]
                and p.return_value = e.process_id
20.    If p is found :
21.       Return p
22.    /* this means e is the staring/root event of a REP */
23.    Return NULL
```

### 2.3.2 Event Linking

All of the generated trace events are linked to form the REP via their parent-child relationship. So, the key of event linking is to identify the parent for each event. Here we describe how the parent is identified for any event of the five categories of event relationships. One event may have multiple parents if more than one rule below applies.

Algorithm 1 describes the event linking for event relationship types of 1~4. The linking for type 4 is lines 3-4, for type 3 is lines 7-9, for type 1 is lines 11-13, and for type 2 is lines 14-21. After Algorithm 1 is executed, the linking for type 5 (data flow of the request processing among threads/processes) is performed.

- *Type 3: Causality of communication between threads/processes.* We insert a unique ID, called MSG_ID, into each transmitted message at the *send* event, and then extract this MSG_ID from the received message at the *recv* event. This MSG_ID is one attribute of the *event linking info*. For any *recv* event, its parent is the *send* event with the same MSG_ID. Unlike vPath which uses TCP connection and socket information to pair *send* event and corresponding *recv* event, our mechanism of message labelling does not rely on specific communication pattern or thread pattern for doing the pairing.

The MSG_ID includes the length information of the message. When a message sent by one *send* event is fragmented and received by multiple *recv* events, or messages sent by multiple *send* events are received in one or multiple *recv* events, the length information is leveraged to identify message borders and merge/split the received fragments into original messages transmitted at *send* events (details of implementation are in Section 3). As a result, one *send* event maps to one *recv* event with the same MSG_ID.

- *Type 1: Temporal successiveness within the same thread.* The *thread info, process info, node info* of events are exploited to find those events generated within the same thread of the given event. Then *local time info* is exploited to identify which event has the time immediately ahead of the given event, i.e. the parent event.

One challenge is to exclude those events that do not participate processing of the current request. For instance, in the thread pool pattern a thread had been created long before the current request was initiated, and those events before the current request should be excluded. In another example, a thread may be serving multiple requests concurrently, and we should exclude those events of processing other requests.

We define a unique ID, MSG_CTX_ID, to aid maintaining the request's context and excluding irrelevant events. The intuition behind the MSG_CTX_ID concept is to group those local events that are related to the same individual message. As the initial service/job request received by the computing platform is a message, the grouped events by this message's context are related to processing of this request. If one of the grouped events is a *send* event, its corresponding *recv* event is identified by the MSG_ID, and the message of this *recv* event creates a new message context with a new MSG_CTX_ID. Therefore, the basic idea is to leverage the cascading use of MSG_CTX_ID and MSG_ID for grouping all events related to the original request. MSG_CTX_ID is also an attribute of *event linking info*.

More formally, MSG_CTX_ID is defined as a unique identifier of the continuous computation from the sending or receiving of one message (event A) to the sending or receiving of the next message (event B). All events between event A and event B along the request execution have the same MSG_CTX_ID. If event A is a *recv* event, a new MSG_CTX_ID is created and marked for both event A and those events after A (and before B); if event A is a *send* event, the event A is still marked with the old MSG_CTX_ID for purpose of event stitching, and a new MSG_CTX_ID is created for those events after A (and before B). The MSG_CTX_ID of event B is similarly assigned as there is a next message after B.

Fig. 3 gives an example of MSG_ID and MSG_CTX_ID propagation along events during request processing. In this figure (and also in our implementation), the events following a *send* event has the MSG_CTX_ID value same as the *send* event's MSG_ID to reduce overhead of generating unique IDs (the *fork* event has id1 as MSG_CTX_ID). Note that the MSG_CTX_ID may propagate across process/thread creation, as shown in Host 1's bottom *read* event which is the first event in a newly-forked process.

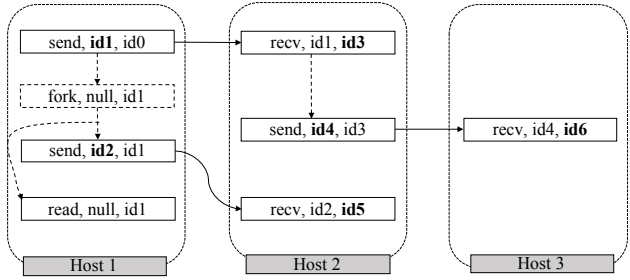

Fig. 2. An example of MSG_ID and MSG_CTX_ID propagation among events (each tuple in the figure is <*call ID, MSG_ID, MSG_CTX_ID*>, a simplified version of the event)

- *Type 2: Causality of thread creation and process creation.* The thread/process-creation event's *call info* has the ID of the newly created thread/process in the call's return value. So if a given event is the first event of a thread/process, we search those thread/process-creation events and identify its parent. MSG_ID and MSG_CTX_ID are used to exclude irrelevant events as well.

- *Type 4: Causality of synchronization among threads.* For each of the synchronization mechanisms such as process waitpid, thread join, signal, etc., we define its specific causality identification method. For example, *waitpid()* blocks the calling process A (the current process) and waits for the specified process B to terminate or change state (stopped or resumed or signaled). Then process B's last event before B's termination or state change is the parent of process A's event that is immediately after the *waitpid()* in the control flow order. The *waitpid()* event of process A is also the parent of this process A's event. Similarly, specific identifications of event causal relationships associated with thread joining, signals, lock/unlock and other IPC mechanisms are defined correspondingly based on their semantics in manipulating execution behavior.

- *Type 5: Causality via data dependency between threads/processes.* General accurate analysis of data dependency relationships along individual traces transparently is a difficult problem to tackle. This paper mainly addresses the construction of REP using control-flow relationships rather than data-flow relationships, as handling data dependency relationships is a radically different topic than handling control-flow relationships. However, we are still able to transparently identify data dependency relationships in these scenarios in certain situations, by leveraging certain IDs which are typically available in many standard queue protocols and middleware. Such IDs include message ID, job ID, session ID, etc. For example, in Apache ActiveMQ, there is a unique message ID in every message header and every job also has a job ID when submitting and executing a job in Hadoop. The data writing and reading events can be correlated by these data IDs. So, in our implemented REPTrace tool we identify the relationships of type 5 in only these situations.

If we do not address the data-induced causality between threads/processes (e.g. message queues) during event linking, the achieved REP is not a single complete path, but comprised of several path segments. In our experiments REPTrace reports 6 paths for a single job's processing when we do not consider such causalities, because job queues or message queues are used in Hadoop. After we leverage the Hadoop job ID and RPC ID to address this type of causality, REPTrace produces one execution path for one job request, as we expected.

### 2.3.3 REP Representation

As an event may have multiple parents, the formed REP via event linking is represented as a graph. Because each linking represents a local temporal successiveness or a causality which is always uni-directed, there is no cycle in the formed REP. So the REP is actually a directed acyclic graph (DAG).

A REP is built for one request's processing. Each node of the DAG represents an event and each edge denotes an event relationship. Each node has zero or a number of child nodes. All the generated events during processing of the request, from the request entering the distributed system until the system replying final results for the request, form the DAG to represent the complete REP.

There might be inconsistency among reads/writes of the same variable/memory location in Type 5 of event relationship (see Section 2.2) if the traced software has unresolved race conditions. If such an inconsistency occurs the REP graph might or might not contain a cycle (we are not sure). But we do not see any cycle in our experiments, maybe because both the probability of activating race condition bugs of the traced softwares (they are widely used mature softwares), and the probability of such inconsistency resulting in a cycle in the REP, are quite low. The linkings of the other four types of event relationships involve either local information only (e.g. local timestamp within a thread) or matching of certain values (process id, thread id, MSG_ID, etc.), and do not incur any inconsistency that may result in a cycle in the REP.

## 3 SYSTEM DESIGN

REPTrace consists of a REPAgent (and an optional Local Generator) on each host of the traced distributed system, as well as a Central Generator (see Fig. 3).

REPAgent intercepts LIBC calls of the host via the LD_PRELOAD mechanism, and emits events to the Central Generator. Our current implementation intercepts 28 LIBC calls in total and emits 28 different types of events, including those examples listed in Table 1. The Central Generator collects these events and runs the event linking in an asynchronous mode (so the performance impact to the request processing is minimized). Then Central Generator stores the complete REP of every request in the repository. If further performance impact minimization is required, an optional Local Generator can buffer REPAgent's events to reduce overhead incurred by the communication to the Central Generator. Then the Local Generator can transmit the trace events to the Central Generator when the system has light load. Due to page limitation, we do not discuss details of this performance optimization.

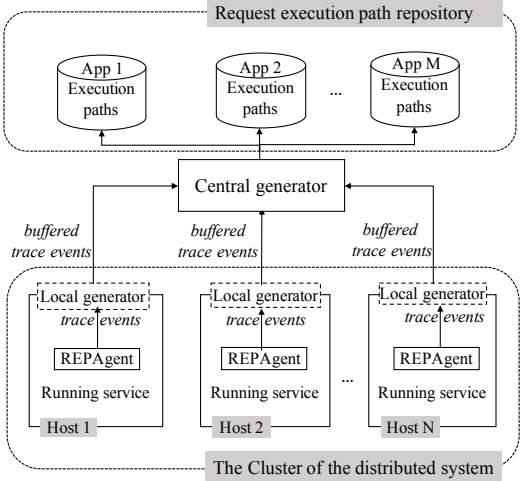

Fig. 3. System architecture of REPTrace

**REPAgent Behavior.** On intercepting a message-sending call, REPAgent a) creates a new MSG_ID and adds the MSG_ID at the beginning of the message, b) invokes the original call to transmit the modified message, c) creates a new MSG_CTX_ID and performs relevant bookkeeping, d) prepares an event and emits it to the Central Generator.

On intercepting a message-receiving call, REPAgent a) invokes the original call to get the message, b) extracts the MSG_ID from the message, c) creates a new MSG_CTX_ID and performs relevant bookkeeping, d) prepares an event and emits it to the Central Generator, and e) returns the message to the upper level of the system.

On interception of other calls, REPAgent performs MSG_CTX_ID bookkeeping, prepares an event and emits it to the Central Generator besides invocation of the original call.

MSG_ID and MSG_CTX_ID are of the same format: a 2-byte starting delimiter, an 8-byte length indicator, a unique-value identifier (UUID) and finally a 2-byte ending delimiter. The length indicator in MSG_ID denotes the length of the original message before modification, and is unused in MSG_CTX_ID.

Each thread has a current MSG_CTX_ID. REPAgent maintains a table of the current MSG_CTX_IDs for all involved threads in the host. On intercepting a message-sending or message-receiving call, the thread's current MSG_CTX_ID is updated with the newly created MSG_CTX_ID. When an event (not a *recv* event) is generated, the event's MSG_CTX_ID is assigned as the thread's current MSG_CTX_ID value (a current MSG_CTX_ID is created for the thread if the ID does not exist yet). Note that the MSG_CTX_ID may propagate across process/thread creation (as Fig. 2 illustrates); therefore, multiple threads' current MSG_CTX_IDs can have same value.

**Handling TCP Message Fragmentation.** TCP messages may be fragmented by the TCP library, and one transmitted TCP message may be received by multiple *recv()* calls. There are also cases when messages transmitted by multiple *send()* calls are received in one *recv()* call (or multiple *recv()* calls without one-to-one mapping between the *send()* calls and *recv()* calls). REPAgent handles the fragmentation during interception of the *recv()* calls. It first reads in the MSG_ID (a fixed number of bytes) and extracts the length information from the MSG_ID. It then reads in multiple messages until the specified length of bytes are received, and returns the merged bytes, i.e. the original message, to the upper application. In the same way, REPAgent is able to split/merge received messages and reconstruct the original messages. Events are properly generated such that one *recv* event maps to one *send* event with the same MSG_ID.

## 4 REPTRACE-BASED ANOMALY DETECTION

REPTrace can be exploited for anomaly detection of the distributed platform behavior. The REptrace-based Anomaly Detection method, READ as the acronym, is composed of two stages: training stage and detection stage. The basic idea is that, a couple of finite state automatons (FSAs) are constructed from the REP during the training, and then the runtime behavior of a request processing is compared against the FSAs to detect anomalies. Though use of FSAs for anomaly detection is not new (Kewei's work [27] is an example), an FSA built from individual request processing's complete end-to-end execution path among a distributed platform, is completely new. Hence, the anomaly detection approach based on this specific kind of FSA is novel.

### 4.1 Training Stage

In this stage we run a number of requests of the same type concurrently on the distributed platform, collect the REP for each of the requests, generate a Finite State Automaton (FSA) for each collected REP, and then merge these per-path FSAs for all the requests of this type, into two aggregated FSAs. The identifications of the parent-child relationships along the execution link all the events into a DAG starting with each single request as the root event.

By request type we mean the function of a service which is requested to be performed. For instances, provisioning of a VM, provisioning of a container, shutdown of a VM, migration of a VM are different types of requests supported by a cloud platform. In a platform with REST API exposed, usually different request types have different URLs and different request modes; in an example command "*curl -ku user:passwd -X POST https://ip:port/service-name/api/v1/the_url -d @jsonobj*", the_url and the request mode POST determine a request type. In a job processing platform like Hadoop, different applications are regarded as different types of job requests.

The two aggregated FSAs will be employed to detect anomaly of the execution of this type of requests in real workload during the detection stage. We repeat this training for all types of requests of interest supported by the distributed platform, so execution anomalies in processing these types of requests can be detected. We aggregate FSAs for each request type rather than for all request types because the processing logics of different types of requests vary a lot (e.g. a VM-provisioning request is completely different from a VM-shutdown request), building a big FSA to tolerate their variations while still provide good detection accuracy has unnecessary complexities, and the accuracy may be sacrificed, too. As the request type information is known at the beginning of the request processing, it is straightforward to do anomaly detection for each request type.

### 4.1.1 Generating a Per-path FSA

Fig. 4 illustrates the process of generating a per-path FSA from one request's REP. The process has the following steps.

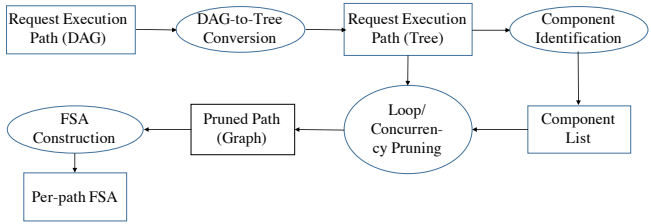

Fig. 4. The Process of Generating a Per-path FSA

### A. DAG-to-Tree Conversion

REP is represented as a DAG because certain events have multiple parents when two or more rules in Section 2.3.2 apply. This multi-parent circumstance brings about computational complexity to the following steps of component identification and path pruning. We convert a DAG into a tree to facilitate these two steps. The conversion is not essential for the loop/concurrency pruning, but mainly for efficiency of the pruning algorithms. Actually, in our experiments less than 1% of event relationshiops are removed during the conversion.

The conversion is done by selecting only one parent relationship for any event with multiple parents and removing the other parent relationships. We propose a subjective prioritization in deciding which rule (in Section 2.3.2) to observe when competing with the other rules, as below (from highest priority to lowest):

*Rule on type 3 > Rule on type 5 > Rule on type 4 > Rule on type 2 > Rule on type 1.*

The same prioritization is observed for all cases; so, the arbitration is deterministic. We select this arbitration principle based on our understanding of typical impacts of different scenarios on request processing behavior.

### B. Component Identification

This operation module identifies the list of components of the distributed system in processing the given request. Component identification is done because a component may have multiple processes/threads, and for human understanding of a system it is highly preferred to understand interactions at the component level rather than the process/thread level. The event tree of the REP is first converted to a process tree (or forest), and then the components are identified from the process tree. The details of the component identification are available in [25].

### C. Loop/Concurrency Pruning

There are lots of loops and concurrencies in processing of a service request. Without dealing with these iterative and concurrent substructures in the REPs, the false positive rate of an anomaly detection method based on the precise REPs will be high due to the variable iteration and concurrency number. Therefore, iteration and concurrency should be identified and pruned. After pruning, the identical iteration and concurrency patterns of events are represented by only one copy for the further analysis, which hugely reduces the number of events and simplifies further processing while does not lose the information of how many times the patterns are repeated (e.g. the iteration number of an iterative substructure is recorded during our processing). The algorithm of the pruning is available in [25].

### D. FSA Construction

An FSA is constructed directly from the pruned REP: each node of the pruned path (i.e. consolidated events) changes into an edge (transition) of the FSA, and each edge of the pruned path changes into a node of the FSA (state between two groups of consolidated events). A state *St0* is added to the FSA before the first transition (i.e. root event) as the start state. Fig. 5 shows an example fragment of a constructed FSA from the pruned REP. An FSA state with out-degree of 2 or more indicates either a branching scenario (conditional statement or loop, e.g. *St4*) or a concurrency (e.g. *St5*). During the loop/concurrency pruning we have information of whether a path fragment is a loop or concurrency. Here we mark the concurrency-caused branching states (*St5* and *St12*) as concurrency points in the built FSA.

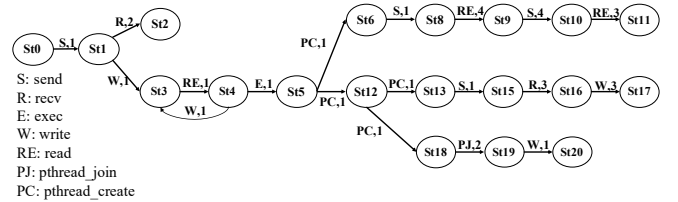

Fig. 5. Example fragment of a built FSA

Note that after we transform a REP into a per-request FSA, the FSA already contains cycles. An FSA with cycles can match infinite number of different execution paths. Though requests may have different input, we observe that the requests of same type with different input usually lead to different numbers of loops or repetitions of same event patterns, and such loops/repetitions are handled by cycles in FSA when we match the execution path against the FSA. Request executions with corner cases of input may be falsely alarmed and result in false positives, and our experimental results report the false positives (Section 5.3).

### 4.1.2 Generating aggregated FSAs for a request type

After building an FSA for each REP, all FSAs for executions of the same type of request are combined into two aggregated FSAs, a core FSA and a full FSA, for anomaly detection.

We observed that normal execution paths of the same type of requests may be slightly different due to various causes such as retries after failures, built-in non-determinism in the distributed platform, etc. Therefore, using the FSA from only one execution path for anomaly detection suffers from a large number of false alarms. We address this issue by combining different FSAs of a request type to generate aggregated FSAs.

The core FSA is the intersection of all the per-path FSAs, and the full FSA is the union of all the per-path FSAs. Thus, the core FSA represents the common execution path fragments/state transitions for a request type, and the full FSA represents all the possible execution fragments/state transitions for all REPs of a request type. The combining process is illustrated in Algorithm 2. The *transition_path(S)* function returns the transition sequence from a per-path FSA's initial state $St0$ to a given state $S$.

Algorithm 2 Combine per-path FSAs
Input: $F_s$, a set of per-path FSAs
Output: the core FSA $F_{core}$ and the full FSA $F_{full}$
1. initialize a hashtable **path_count**   // count how many FSAs a path is in
2. **For** each per-path FSA $F$ in $F_s$ :
3.   **For** each state $S$ in $F$ :
4.     state_path $SP$ = **transition_path**($S$)   // a sequence from $St0$ to $S$ in $F$
5.     **If** $SP$ **not in path_count** :
6.       add $SP$ to **path_count**
7.       **path_count**($SP$) = 1
8.     **else**
9.       **path_count**($SP$) + =1
10. **For** each state_path $SP$ in **path_count** :
11.   **If path_count**($SP$)= **size**($F_s$) :   // the path is in all per-path FSAs
12.     Merge($F_{core}$, $SP$)
13.     Merge($F_{full}$, $SP$)
14.   **else**
15.     merge($F_{full}$, $SP$)
16. **Return** $F_{core}$, $F_{full}$
17. Merge($F$, $SP$):
18.   Match $SP$ against $F$ starting from $St0$
19.   **If** $SP$ **not in** $F$, i.e. a branching state $S_{br}$ is identified :
20.     link the rest sequence of $SP$ into $F$ at $S_{br}$

**Annotating aggregated FSAs with time information.** We enhance the two aggregated FSAs by annotating time information to each transition of the two FSAs. The annotated time information is the average of the time spent on those REP trace events which were consolidated into this FSA transition during the step of *Loop/Concurrency Pruning* (Section 4.1.1 C) and the step of *Generating aggregated FSAs* above in this Section (recall that each REP trace event has the *timestamp*). The time annotation can be leveraged to detect performance anomalies.

## 4.2 Detection Stage
The aggregated FSAs are applied to detect anomalies of request executions. The detection process starts when a request is first received by the distributed platform, e.g. at the portal of a cloud platform, at the first REST API component of a request processing system, or at the Hadoop component that receives a user-initiated job request. Specifically, the request type is identified, the corresponding aggregated FSAs for the request type are selected, and the anomaly detection is performed against the aggregated FSAs.

### 4.2.1 Identifying the request type
As the REPTrace intercepts the network messages to and from all components of the target distributed platform, we add platform-specific handling in the interception of the component that receives external requests from clients. The platform-specific handling parses the external requests, identifies the request type, and sends the request type to the Central Generator (Fig. 2). Then the aggregated FSAs for this identified type of request are selected for anomaly detection in the following step.

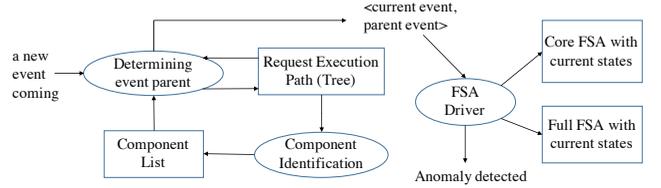

Fig. 6. Detection of execution anomaly against aggregated FSAs

### 4.2.2 Detecting Anomaly of Request Execution
Fig. 6 shows the overall procedure of detecting execution anomaly against the two aggregated FSAs. Three modules run in parallel during the detection, and the modules maintain four data structures as the execution of the request goes on: a tree-form of the REP, a component list, the core FSA and the full FSA with the current states of the request execution marked in the two FSAs. Here we describe the behavior of the three modules.

**A. Determining event parent**

When a new REPTrace event is reported to the Central Generator, this module performs these two tasks:

i) identifies the parent event of the incoming trace event based on the current REP and links the incoming event into this REP. The arbitration prioritization described in Section 4.1.1 A is applied for multi-parent cases. Therefore, the result REP is a tree. The time complexity of this step is $O(n)$ where n is the current number of events in this REP.

ii) adds the *component_id* attribute to the new event based on the information of the component list. Then the event and its parent event are passed to the FSA driver.

**B. Component Identification**

After the REP is updated upon the new event, this module processes the REP and updates the component list as described in Section 4.1.1 B.

**C. FSA Driver**

As the request may be executed by multiple components of the distributed platform in parallel at a time, multiple events may be produced concurrently. This means the current state of the distributed platform in processing the request is actually captured as a set of active FSA states in both the core FSA and the full FSA (i.e., one active FSA state indicates the current state of one thread that is processing the request at this time), similar to tokens in a classical Petri-Net.

Recall that a trace event of REPTrace maps into an FSA edge during FSA construction. So, when a pair of *<current event, parent event>* arrives, the FSA driver triggers state transition in each FSA by i) identifying the source FSA state which is one of the active FSA states and is also between two FSA edges matching the incoming events (match of *ftype* and *component_id*), and ii) transitioning the activeness of the source state to the next FSA state according to the current event.

*Handling concurrency.* A *pthread_create/fork/vfork* trace event generates two transitions and one active FSA state results in two active states as one more thread/process is created. Moreover, as this is online anomaly detection, an identified concurrency point of the FSAs (see Section 4.1.1 D) stays active after it is activated, to deal with the scenarios when events of concurrent operations arrive with large time gaps.

*Detecting Functional Anomalies.* In one of the following conditions, an anomaly is detected.

- The <*parent event, current event*> pair cannot trigger a transition in the full FSA, i.e., the match fails and the source FSA state cannot be identified.

- At least one transition of the core FSA is not traversed by the execution when the request processing finishes. As events arrive frequently during the execution, when no event of the request processing arrives within a threshold period of time (5 seconds in our experiments), it is considered the request processing completed.

When an anomaly is detected, the detailed information of the corresponding transition/event, the collected REP and the component list is available for further investigation and diagnosis.

*Detecting Performance Anomalies.* The aggregated FSAs are annotated with time information. A performance anomaly is detected when the measured time in the current event is longer than the annotated time information of the transition by a given threshold percentage (100% used in our experiments as a subjective selection).

## 5 EXPERIMENT AND EVALUATION

We implemented a prototype of REPTrace on Linux distributions and the source code can be found on GitHub [19]. Our current implementation intercepts 28 LIBC calls via the LD_PRELOAD mechanism on Linux, including the POSIX library calls on network operations and process/thread operations as listed in Table 1.

**Experiment Setup.** The scenarios evaluated include:

a) comparison of the tracings of the WordCount application on Hadoop (including Yarn, MapReduce and HDFS) by REPTrace and vPath, for evaluating the correctness and completeness of the REPTrace;

b) overhead analysis for the tracings of applications by REPTrace, on Hadoop, Spark, Tensorflow (for image classification using SVM) and Angel [20];

c) REPTrace-based anomaly detection for four typical applications on Hadoop (i.e. four request types, as a different application is regarded as a different request type for the Hadoop job processing system): Kmeans, WordCount, Grep and TopN.

The workloads in our experiments include WordCount, Grep, Kmeans, TopN, Teragen and Terasort on Hadoop, WordCount on Spark, Logistic Regression on Angel and SVM on Tensorflow. These workloads are classical and common workloads for Hadoop, Spark, TensorFlow and Angel, respectively. For each application (request type) we run 10 jobs (requests) for overhead evaluation, 20 jobs for the training stage of the anomaly detection, and 28 normal jobs, 48 fault injected jobs for the detection stage. The experiments were conducted on 2 KVM virtual machines on a Dell R730 server. Each VM has 4 cores of Intel Xeon E5-2603, 4GB main memory and Ubuntu 16.04.

### 5.1 Request Execution Path Analysis

We conducted experiments to compare vPath and REPTrace on tracing of the WordCount application on Hadoop with a 500M input file. We got the vPath source code from the authors of the vPath.

vPath averagely generates 28,908 trace events while REPTrace averagely generates 39,555 trace events. We investigate a subset of the events and confirm the events and their relationships captured by vPath are all captured by REPTrace. However, vPath produces 1,153 execution path fragments, while REPTrace produces 6 execution path fragments for one job execution before the handling of data dependency relationships in REPTrace is invoked. vPath produces much larger number of fragments because it only supports a constrained communication/thread pattern.

After we invoke REPTrace's handling of Hadoop job ID/RPC ID (a plugin of REPTrace for supporting Hadoop's data dependency relationships), the 6 execution path fragments are linked together to form a single complete REP for the job request. We then looked into the events of several REP paths, and did not find any error in inspected relationships of the paths (via manual inspection of a number of generated REPs and relevant source code), i.e. all inspected ones reflect the real event relationships as expected and as specified in Section 2.3.2.

#### 5.1.1 Case Analysis

vPath reports more than one thousand path fragments for processing of one request because Hadoop RPC and certain communication mechanisms in Hadoop do not support vPath's assumed pattern, which is described in Section 1.

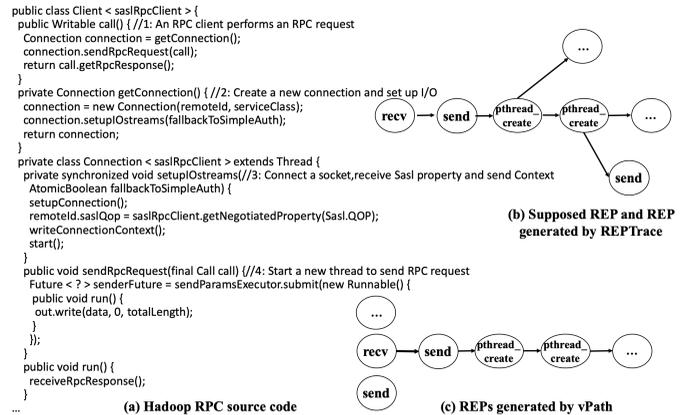

Fig. 7. Hadoop RPC and its partial REP

Fig. 7(a) shows a simplified Hadoop RPC client implementation. An RPC client calls *call()* to send an RPC request to an RPC server, which creates a new connection and sets up I/O streams for this connection. In *setupIOstreams()*, the RPC client calls *getNegotiatedProperty()* to receive necessary authentication information from the connection. Then the RPC client launches a thread to receive RPC responses and another thread to send the RPC request via this connection. The expected REP fragmemt for this scenario (starting from receiving the negotiation information), which is the same as the REP fragment generated by REPTrace, is shown in Fig. 7(b). The REP fragments produced by vPath are shown in Fig. 7(c). The events of the same request are identified as three path fragments by vPath. Here are two cases when vPath fail to link events correctly.

*Case 1.* vPath assumes that after receiving a request X (*recv-request-X* event), the same thread will emit subordinate request Y (*send-request-Y*) to another component, block itself, receive the reply from that component, and then reply to request X (*send-reply-X*). But this assumption fails to hold in the scenario above where the *recv-request-X* event and *send-request-Y* event occur at different threads. This request-reply pattern is widely used in Hadoop, especially in Hadoop RPC which is used by Hadoop components to communicate with each other.

So vPath fails to capture the relationships between one thread's *recv* event and the *send* events of the thread's child threads. This thread pattern is handled by REPTrace propagating MSG_CTX_ID across thread/process creation, as discussed in Section 2.3.2 *Type 1* and *Type 2* parts.

*Case 2.* vPath also assumes synchronous network communication that a thread sends a request message over a TCP connection and then blocks until the corresponding reply message comes back over the same TCP connection. However, as shown in Fig. 7(a), RPC client starts a thread to perform *sendRpcRequest()* and another thread to perform *receiveRpcResponse()*. The *send* and *recv* are asynchronous and are not done within same thread.

Therefore, vPath fails to connect these two events and produces two unrelated path fragments. REPTrace doesn't depend on the synchronous communication pattern. Instead, REPTrace employs MSG_ID and MSG_CTX_ID cascadingly to maintain request context across components, and is able to identify the asynchronous *send* and *recv* in two threads are in the same request processing, and link them into the same path.

### 5.2 Overhead Analysis

There are two types of overhead by using REPTrace to trace a distributed system: increase of the request latency and increase of network traffic due to the transmission of extra information such as MSG_ID. Specifically,

- *Latency Overhead = (average latency with REPTrace – average latency without REPTrace)/average latency without REPTrace*
- *Traffic Overhead = size of MSG_ID/average size of messages*

Each application of each system is run 10 times for the overhead analysis. Table 2 presents the results.

Table 2 shows that REPTrace captures REPs for different distributed platforms with a 4.5% average latency overhead and a 1.8% average traffic overhead. A smaller average size of messages leads to a higher traffic overhead. Tensorflow's traffic is mainly for the parameter synchronization between the worker and the parameter server to update the parameters of machine learning models; it has much smaller message sizes compared to other distributed systems in the table. Consequently, Tensorflow's traffic overhead is higher.

TABLE 2: Results of the overhead analysis

| Systems | Applications | Latency Overhead | Traffic Overhead |
|---|---|---|---|
| Hadoop | WordCount | 6.0% | 0.8% |
| | Teragen | 4.6% | 0.2% |
| | Terasort | 4.0% | 0.3% |
| Spark | WordCount | 4.4% | 0.3% |
| Angel | Logistic Regression | 4.8% | 0.9% |
| Tensorflow | SVM Classification | 3.1% | 8.2% |
| Average | - | 4.5% | 1.8% |

### 5.3 Functional Anomaly Detection

During the training stage we run 20 jobs for each of the 4 applications on Hadoop (see the experiment setup at beginning of Section 5 for details), with different input files of different sizes. The numbers of trace events vary with the input size. For a WordCount job with 500M input file, the average number of trace events in the reported REPs is 39,550 and the built full FSA has 16,297 states.

**Fault injection.** We use fault injection to evaluate the detection performance. We inject 6 categories of failures: Output path collision, Namenode crash, Datanode crash, Resource Manager crash, NodeManager crash and Runjar crash. For the output path collision, we started a Hadoop job request and then had another task to lock the output path of the request processing. For other categories of failures, we killed the corresponding component process after we confirmed the process is running and the processing of the request already started. All of the failures are activated and manifested in our experiments.

**Results.** For each application, we ran 28 normal jobs and 48 jobs with fault injection (8 jobs for each of the 6 failure categories). Therefore, there are 304 jobs in total, 112 without fault injection and 192 with fault injection. We use all of the 20 REPs constructed during training to train the anomaly detection; the results are summarized in Table 3.

TABLE 3: Functional anomaly detection results of four applications (FSA-20)

| Application | Precision | Recall | F1-measure |
|---|---|---|---|
| Grep | 0.98 | 0.85 | 0.91 |
| Kmeans | 0.92 | 1 | 0.96 |
| TopN | 0.84 | 1 | 0.91 |
| WordCount | 1 | 1 | 1 |
| Average | 0.93 | 0.96 | 0.94 |

We also evaluate the impact of the number of training REPs and event pruning on the detection performance. The anomaly detection results for four models are listed in Table 4.

- *FSA* in the table refers to an FSA model directly built from one REP for each application without pruning loop or concurrency. As only one REP path is used, there is only one FSA constructed and there is no core FSA or full FSA.
- *eFSA* refers to an FSA model built from one REP for each application with pruning of loop and concurrency. There is only one FSA and no core FSA or full FSA.
- *FSA-n* refers to a core FSA and a full FSA built from n training REPs (n<=20 as 20 jobs are executed in training) for each application, with loop and concurrency pruning.

YONG ET AL.: TRANSPARENTLY CAPTURING EXECUTION PATH OF SERVICE REQUEST PROCESSING FOR ANOMALY DETECTION 11TABLE 4: FUNCTIONAL ANOMALY DETECTION RESULTS WITH DIFFERENT MODELS

| Model | Precision | Recall | F1-measure |
|---|---|---|---|
| *FSA* | 0.034 | 1 | 0.07 |
| *eFSA* | 0.426 | 1 | 0.60 |
| *FSA-10* | 0.837 | 1 | 0.91 |
| *FSA-20* | 0.930 | 0.96 | 0.94 |

Table 4 shows that without pruning of loop and concurrency, *FSA* has very low precision, which indicates huge false positives because normal jobs' execution paths differ with each other due to iteration, concurrency, fault tolerance or built-in non-determinism. *eFSA* improves the precision from *FSA* and reduces the false postives after pruning loop and concurrency, but the false positives are still pretty high. *FSA-10* and *FSA-20*, by combining the core FSA model and the full FSA model from multiple normal REPs, alleviate the false positive issue to an acceptable level. With FSA-20, the average functional anomaly detection precision is 93% and the recall is 96%, which shows that the REPTrace-based Anomaly Detection (READ) has a strong capability to detect functional anomalies.

## 5.4 Performance Anomaly Detection

To evaluate the performance anomaly detection, three types of performance faults are injected to the running Hadoop systems, as Table 5 lists, while jobs are running. We ran 9 normal jobs and 9 jobs with fault injection for each of the 4 applications, 3 jobs for each of the three fault types. Therefore, there are 72 jobs in total, 36 with fault injection and 36 without. On average, READ achieves a 74% precision and a 74% recall for performance anomaly detection. We analyzed the factors that limit the precision and recall of performance anomaly detection. The results show that more advanced mechanism than our simple threshold-based detection should be explored in our future work.

TABLE 5: PERFORMANCE ANOAMALY DETECTION RESULTS OF INJECTED PERFORMANCE FAULTS

| Performance Fault | Precision | Recall | F1-measure |
|---|---|---|---|
| CPU burning | 0.70 | 0.78 | 0.74 |
| High network latency | 0.88 | 0.78 | 0.83 |
| Full disks | 0.67 | 0.67 | 0.67 |
| Average | 0.74 | 0.74 | 0.74 |

## 6 RELATED WORK

### 6.1 End-to-end Tracing Systems

End-to-end tracing has been a fertile research area and there exist various implementations in both academia [1], [3], [7], [11], [12], [13], [15], [18] and industry [6], [16], [17]. Pinpoint [11] instruments J2EE to accurately correlate the network communication events by adding request identifier in HTTP request headers and correlate intra-unit events by automatically propagating request identifiers with thread local variables of a thread. Stardust [15] is an end-to-end tracing system designed for specific distributed storage system, which requires manual modification of the storage system to generate the request execution path. Xtrace [13] mainly focuses on extending network protocols to support request identifier propagation to generate task trees. Dapper [6] is Google's internal distributed tracing system serving various clusters. Dapper instruments internal homogeneous control flow/RPC libraries to generate request execution path without source code modification of upper applications. HTrace [17] is a Dapper-like end-to-end tracing framework, which has been integrated with HDFS and HBase.

Stardust requires manual modification of the distributed storage system. Pinpoint, Dapper, Pivot and Canopy deal with specific middleware or vendor custom libraries only. HTrace requires instrumentation to the system to enable tracing. Certain hardware and application support is demanded for Xtrace to work.

A project close to our work is vPath [7]. This project made a thrust to provide a generic solution of tracing distributed systems/applications by monitoring thread and network activities. vPath only deals with TCP-based system/applications and assumes certain communication styles and thread patterns. For example, vPath assumes that i) any thread A, after it sends a message to thread B, it must wait until receiving the response from B; A should not send or receive another message during the wait; ii) any component P, after it receives a message from an upstream component Q, all subordinate messages from P to downstream components must be sent by the same thread of P which receives the Q's message. These are very strict assumptions of communication styles and thread patterns, and do not hold in many modern distributed systems. As a result, vPath generates many execution path fragments rather than an accurate complete execution path for one request in many scenarios, as demonstrated in our evaluation.

There are also works that apply statistical technologies to logs of the distributed components for correlating log events. They apply statistical analysis [8, 14], code analysis [10], time series analysis [5], or timestamps [4, 9] to speculate the request execution path. Because many events are not exposed by the logs, these technologies cannot capture the accurate complete execution path of fine-grained events for individual request and just try to probabilistically matching logs events with statistical inference or identify certain statistical patterns.

### 6.2 Anomaly Detection of Distributed Systems

Previous research works make use of different types of runtime data to perform anomaly detection in distributed system, such as log, monitoring data or tracing data.

Fu [21] leverages log data and assumes that there is a thread ID or request ID for every log event to classify all the logs into different log sequence according to their ID information and timestamp. An FSA (Finite State Automaton) is built for each component of distributed system based on the log sequences. Then the built FSAs are used to perform workflow and performance anomaly detection. Due to its sequence model, it fails to depict the concurrency in a request and is limited to a single component without correlating the log events from other components or hosts. Wang [22] uses monitoring data to build anomaly detection model based on different monitoring metric. But it can only detetect the overall performance anomaly and is not applicable to functional anomaly and request execution performance anomalies. Sambasivan [23] diagnose both workflow and performance anomaly by comparing execution paths generated by Stardust. It identifies parallel substructures of request execution path but fails to find the iterative substructure of a request execution path. So, there will be a lot of false alarms when applying it to a distributed system with frequent iteration. We are not the first one to exploit FSA to detection anomiles. KeWei [27] proposed an FSA-based problem detection approach. This approach parses log events, uses <*component, log_point*> tuples as the finite states, uses temporal order as the transitions for states, to build a FSA for all possible service requests. Temporal order is not accurate to capture the transitions between these states to build an accurate FSA. Meanwhile, its detection is based on the historical states transition frequency between any two states, which doesn't consider the context of the current state.

## 7 CONCLUSION

In this paper, we propose REPTrace, a generic end-to-end tracing methodology, which automatically generates the complete execution path of service/job requests for a variety of distributed platforms in a transparent way. We analyzed possible request execution scenarios during distributed platforms' request processing and corresponding causal relationships, and introduced how REPTrace captures the accurate causal relationships in these scenarios. We conducted experiments to compare with vPath. Our results show that REPTrace has much better performance in terms of generality and accuracy because REPTrace is not limited to certain network communications styles or thread patterns as vPath is. In our experiments REPTrace has a low execution latency overhead (4.5%) and low network traffic overhead (1.8%). We also devise an anomaly detection mechanism based on REPTrace to detect anomalies of request execution. The experiments on Hadoop show that functional anomalies are detected with 93% precison and 96% recall by leveraging the REPs.